\documentclass{article}[11]

\usepackage{graphics,latexsym}
\usepackage{graphicx}
\usepackage{amsmath, amssymb,amsthm}

\usepackage{color}

\begin{document}

\title{Charging Capacitors According to Maxwell's
Equations: Impossible}

\author{Daniele Funaro$^*$ }
\date{}
\maketitle

\centerline{$^*$\small Dipartimento di Fisica, Informatica e Matematica }
\centerline{\small Universit\`a di Modena e Reggio Emilia, Via Campi 213/B, 41125
Modena (Italy)}
\medskip

\begin{abstract} {\small The charge of an ideal parallel
capacitor leads to the resolution
of the wave equation for the electric field with prescribed
initial conditions and boundary constraints. Independently of the
capacitor's shape and the applied voltage, none of the
corresponding solutions is compatible with the full set of
Maxwell's equations. The paradoxical situation persists even by weakening
boundary conditions, resulting in the impossibility to describe
a trivial phenomenon such as
the capacitor's charging process, by means of the standard Maxwellian
theory.}
\end{abstract}

\vspace{.2cm}
\par\smallskip


%02.30.Jr Partial differential equations  
%41.20.Jb Electromagnetic wave propagation 
%01.55.+b General physics  

\maketitle

\section{Background}

The subject here are the
standard Maxwell's equations and their inability to handle a lot
of situations that in the common practice are instead considered
trivial. The main criticism is that the system is
overdetermined, i.e., solutions must satisfy too many constraints
without enjoying the necessary degrees of freedom.
The discovery of these inconsistencies was made
about ten years ago, when I started a review process of the theory of
electromagnetism. The underlying motivation was based on some
unsatisfactory marginal aspects. Nevertheless, in the development
of the analysis, these aspects became much more relevant. The
result was a renewed model (see \cite{fun1}, \cite{fun2}) that
strictly includes the solutions to Maxwell's equations, thus
providing the description of a wider range of events. In
particular, non-dissipating compact-support electromagnetic waves,
travelling straightly at the speed of light, are very easily
modelled by the new set of equations. The importance of this fact
is high if we realize that one of the reasons for the split of
physics into the classical and the quantum versions is actually
the impossibility to represent photons via Maxwell's equations in
vacuum. Indeed, an initially localized wave-packet, whose fields
are successively modelled by Maxwell's equations, is soon
destroyed, diffusing all around.
\par\smallskip

It is not my intention to further self-celebrate here the potentiality
of my extension and the possible implications in the study of the
quantum world by a classical approach. For further insight
the reader is referred to \cite{fun4}. 
\par\smallskip

Criticizing Maxwell's equations is dangerous. One is immediately
relegated as heretic. On the other hand, the power of mathematical
reasoning cannot be ignored. After publishing my first report
(\cite{fun1}), I
was contacted by Dr. W. Engelhardt  (see \cite{eng1}, \cite{eng2},
\cite{eng3}). He was puzzled by the excessive number of
constraints that a Maxwellian solution has to satisfy. By trying
to impose all of them one inevitably comes to contradictions.
Usually, engineers follow a certain computational path in order to
come out with solutions mimicking as much as possible reality.
When a reasonable output is obtained they do not feel it is
necessary to check if further restrictions apply. It is like
storing all of $n+1$ objects inside $n$ boxes (one object per
box). There are several paths one can follow, but none is going to
be resolutive. A dirty trick is to hide the last object (whatever
it is) and show to the public one of the allowed combinations.
\par\smallskip

Due to its linearity, it is easy to prove that the Maxwell's
system must have, under very mild assumptions, at most one solution 
(uniqueness). The work of
Engelhardt shows that there may be different solutions to the same
given problem. How can this happen? In reality, the problem
imposes $n+1$ constraints and turns out to be impossible; however,
by applying $n$ steps of the solution process one can actually get
something meaningful. That ``something'' depends on the constraint
that has been discarded.  Engelhardt writes the electromagnetic
unknowns in terms of the potentials $\Phi$ and ${\bf A}$ and notes
that different conclusions are reached according to the choice of
the {\sl gauge}, notwithstanding that the gauge has no influence on the
expression of Maxwell's equations. The conclusion that there are
different solutions contradicts uniqueness. This nonsense can only
be justified by deducing that none of the solutions proposed is
correct, not because of a mistake in the computation, but because
they result from an incomplete procedure and that a full
resolution does not exists. This observation casts dark clouds on
the Maxwellian theory.
\par\smallskip

In this short note I would like to study in the easiest possible
way a very simple problem: the charge of a capacitor. Solving the
wave equation for the electric field I  get a solution
incompatible with Amp\`ere's law. A similar question was examined
in \cite{eng3} via retarded potentials. There the author points
out inconsistencies between the wave equation and the Faraday's
law. Hence, the analysis of the distribution of an electromagnetic
field inside a capacitor depends on the way the problem is
presented. In the quasi-stationary regime (current flow is
relatively slow) the magnetic contribution is usually neglected.
If higher modes are invoked, the full system of Maxwell's
equations must be involved. In this fashion, according to the
nature of the phenomenon, one picks up the right tool to operate,
consisting of $n$ relations chosen on purpose. Very often the
outcome is convincing. If little troubles emerge they are
attributed to some unavoidable inaccuracy dictated by the limits
of the model. However, the question is deeper: the model itself is
not mathematically correct when taken with all its constraints.
One could find rigorous mathematical outcomes by getting rid of
one constraint at the time, but at this point there is no a unique
theory of electromagnetism.
\par\smallskip

In \cite{eng3}, the limit of the Maxwellian theory is attributed
to the presence of nonhomogeneous terms. In my opinion, as
rigorously analyzed in \cite{fun3}, even homogeneous Maxwell's
equations in vacuum are not trouble free, displaying an extremely
reduced space of solutions. The equations in this case are
affected by an almost total lack of initial displacements
satisfying both the conditions ${\rm div}{\bf E}=0$ and ${\rm
div}{\bf B}=0$ (see my viewpoint in \cite{fun4}, chapter 1).
\par\smallskip

It must be honestly pointed out however that the solution's space of Maxwell's
equations is far from being empty. There are in fact remarkable
situations where the model has mathematical meaning and matches
reality. Solutions seem however to belong to a kind of {\sl meagre
set}. This set looks closed, connected and with zero measure (with
respect to standard topologies in functional spaces), though I
have no rigorous proof of these statements. Common
applications, as the one studied here, belong to the {\sl complementary
set} and cannot be approximated by Maxwellian solutions. My conclusion is that,
despite of the extraordinary achievements of the technological
world, Maxwell's\footnote{Allow me to give my respects to J.C. Maxwell, who never wrote
the equations in the form we are used to and never imagined that his name would have been
involved in such diatribes.} equations are inefficacious in most practical cases, unless
they are accompanied by rough and non well justified mathematical
adaptations.

\par\medskip

\section{Technical preliminaries}

In order to make the problem as simplest as possible let us
suppose we are in vacuum, though this restriction is not
necessary. Maxwell's equations include the Amp\`ere's law:
\begin{equation}\label{fem1}
\frac{\partial {\bf E}}{\partial t}~=~ c^2{\rm curl} {\bf B}
\end{equation}
where we set the current source term equal to zero.
Moreover, we have the Faraday's law of induction:
\begin{equation}\label{fbm1}
\frac{\partial {\bf B}}{\partial t}~=~ -{\rm curl} {\bf E}
\end{equation}
and the two following conditions on the divergence of the fields:
\begin{equation}\label{fde1}
{\rm div}{\bf E} ~=~0
\end{equation}
\begin{equation}\label{fdb1}
{\rm div}{\bf B} ~=~0
\end{equation}
Putting all together, there are six unknowns and
eight equations.
\par\smallskip

Relations (\ref{fde1}) and (\ref{fdb1}) are often taken as an
optional. For example, if one assumes that ${\rm div}{\bf E}=0$
holds at initial time $t=0$, it follows from (\ref{fem1}) that the
divergence of the electric field must be zero at all times. It is
enough to compute the divergence of both terms in equality
(\ref{fem1}) to obtain:
\begin{equation}\label{dfem1}
\frac{\partial ({\rm div}{\bf E})}{\partial t}~=~
c^2~{\rm div}({\rm curl} {\bf B})~=~0  ~~~~~~~~~\forall~t
\end{equation}
The above passage is mathematically correct. However, it is source
of big mistakes. From expression (\ref{dfem1}) we presume that it
is not necessary to check whether equation ${\rm div}{\bf E}=0$ is
maintained during time evolution. Nevertheless, it would be wise
not to be much confident on this fact. We shall demonstrate
through a simple example that ${\rm div}{\bf E}$ can instead
spontaneously assume values different from zero, mining the
foundations of the Amp\`ere's law in vacuum.
\par\smallskip

We assume that our functions are regular, so that they can be
differentiated as many times as needed. It is easy to recover the
following equation regarding the time variation of the
electromagnetic energy:
\begin{equation}\label{poyn}
\frac{1}{2}\frac{\partial}{\partial t} (\vert {\bf E}\vert^2
+c^2\vert {\bf B}\vert^2)
~=~- c^2~{\rm div}({\bf E}\times {\bf B})
\end{equation}
Here ${\bf E}\times {\bf B}$ is the Poynting vector. To get
(\ref{poyn}) one scalarly multiplies (\ref{fem1}) by ${\bf E}$,
(\ref{fbm1}) by ${\bf B}$, and uses notions of standard calculus.
\par\smallskip

The electric field satisfies the wave equation:
\begin{equation}\label{wave}
\frac{\partial^2 {\bf E}}{\partial t^2}~=~ c^2\Delta {\bf E}
\end{equation}
which is obtained by observing that:
\begin{equation}\label{curldel}
\Delta {\bf E}~=~- {\rm curl}({\rm curl}{\bf E})~+~\nabla ({\rm div}{\bf E})
\end{equation}
Therefore, relation (\ref{fde1}) is necessary in order to get
(\ref{wave}). The wave equation also holds for the magnetic field
${\bf B}$.
\par\smallskip

In addition, there are initial conditions and boundary
constraints. The discussion of these is a crucial issue. There are
several ways to impose boundary conditions. Commonly, a list of
possible choices is presented (see, e.g., \cite{jackson}, section
I.5, or \cite{grant}, section 11.6.1). One can then pick up the
ones that better fit the phenomenon to be studied.
\par\smallskip

As in the case of the wave equation (\ref{wave}), there is the
tendency to transform the first-order system (\ref{fem1}),
(\ref{fbm1}) into a second-order one. Playing with second-order
derivatives in the space variables is much easier and the {\sl
well-posedness} of the problem generally follows from properties
of the Laplace operator. In this circumstance, boundary conditions
are naturally derived from a solid theory. It is to be noticed
however that the choice of the constraints for an elliptic
operator is not equivalent to that of a system of hyperbolic
equations, where a preliminary study of the {\sl characteristic
lines} should be done in order to detect which part of the
boundary is actually involved. This analysis looks quite difficult
in the context of Maxwell's equations, where the notions of
characteristic curves and wave-fronts are in most cases not very
clear.
\par\smallskip

For practical applications, the nature of the boundary constraints
comes from physical considerations, sometimes without worrying
about the mathematics. It is not rare to see cases where boundary
conditions are under-determined, and others in which boundary
conditions are over-determined. Nevertheless, this does not seem
to cause any sort of ethical problem. As far as the results are in
agreement with reality there is no reason to suspect the
possibility of spurious solutions or that the entire formulation
is inconsistent.
\par\smallskip

By examining a specific case, let us review what possibilities are
offered regarding equation (\ref{wave}). For a given smooth
bi-dimensional domain $\Omega$, we consider a capacitor where the
two plates, shaped as $\Omega$, are parallel and placed at a
distance $d$. The vertical direction is the $z$-axis and the
plates are situated at the positions $z=0$ and $z=d$. We assume to
work with an {\sl ideal capacitor}. This means that the electric
field stays perpendicular to each plate surface and that the
charge can be uniformly modified on the plates. As initial
condition we impose (capacitor completely uncharged):
\begin{equation}\label{init}
{\bf E}~=~0 ~~~~~~~~  {\bf B}~=~0 ~~~~~~~{\rm at~time~~}t=0
\end{equation}
Since ${\bf B}=0$, one has ${\rm curl} {\bf
B}=0$ for $t=0$. Thanks to (\ref{fem1}), one must have:
\begin{equation}\label{initd}
{\bf E}~=~0 ~~~~~~~~~~~\frac{\partial {\bf E}}{\partial t}~=~0
~~~~~~~~~{\rm at~time~}t=0
\end{equation}
\par\smallskip

The behavior of the electric field on the two plates will be
specified later.
We assume that laterally the capacitor is totally insulated,
that amounts to say that the electric field is orthogonal to the
normal ${\bf n}$ to the boundary $\partial\Omega\times [0,d]$.
By setting ${\bf E}=(E_x,E_y,E_z)$ and ${\bf n}=(n_x,n_y,n_z)$,
we must have $n_z=0$, $n_x^2+n_y^2=1$ and:
\begin{equation}\label{inse}
{\bf E}\cdot{\bf n}~=~ n_xE_x+n_yE_y~=~0 ~~~~~~~~~{\rm on}~\partial\Omega\times
[0,d]~~~~~\forall ~t
\end{equation}
We would like to discuss the uniqueness of the solution of the
vector wave equation. To this end, relation (\ref{inse}), being
just a scalar equality, is not sufficient. In fact,
we are not assigning Dirichlet boundary conditions on both
components $E_x$ and $E_y$, but only a constraint on a linear
combination of them. Something more is required.
\par\smallskip

A good companion for (\ref{inse}) is the boundary relation ${\bf
B}\times {\bf n}=0$, $\forall~t$. This can be differentiated in
time, obtaining $(\partial {\bf B}/\partial t)\times {\bf n}=0$,
$\forall~t$. Recalling (\ref{fbm1}), we can translate the last
relation in terms of the curl of ${\bf E}$. Therefore, a suitable
set of boundary constraints for the lateral surface of the
capacitor is:
\begin{equation}\label{rote}
{\bf E}\cdot{\bf n}~=~0 ~~~~~~~~~~{\rm curl}{\bf E}\times{\bf n}~=~0
~~~~~{\rm on}~\partial\Omega\times [0,d]~~~~~\forall ~t
\end{equation}
Traducing in terms of components, one has:
\begin{equation}\label{rotec}
E_xn_x+E_yn_y=0~~~~~~~~~\frac{\partial E_x}{\partial y}=
\frac{\partial E_y}{\partial x}~~~~~~~~~\frac{\partial E_z}{\partial
x}n_x+\frac{\partial E_z}{\partial  y}n_y=0
\end{equation}
i.e., the right number of constraints. Since ${\bf B}\times {\bf
n}=0$ implies that the magnetic field is orthogonal to the
boundary, it turns out that the Poynting vector ${\bf E}\times
{\bf B}$ is tangential to the surface, that is in agreement with
the fact that the energy cannot escape from the walls (see later
on).
\par\smallskip

We are going to show that the wave equation for the electric
field has unique solution (which is not however a proof for
existence). If there were two distinct solutions, their difference
would satisfy the wave equation with homogeneous data.
Consequently, we impose the initial conditions in (\ref{initd}),
Dirichlet homogeneous conditions on the upper and the lower
plates, and conditions in (\ref{rote}) on the lateral surface.
It is possible to show (see below) that, with these constrictions,
the only admissible case is  ${\bf E}$ identically
zero everywhere at every time.  Therefore, the difference of two
solutions has to vanish identically, and this is against the
hypothesis that they are distinct.
\par\smallskip

The uniqueness of the vector wave equation follows from the
conservation of a suitable energy. By extending the proof given in
\cite{evans}, p. 83, to the vector case, one scalarly multiplies
both members of (\ref{wave}) by $\partial {\bf E}/\partial t$ and
integrates on the whole domain:
\begin{equation}\label{wave2}
\int_{\Omega\times [0,d]} \frac{\partial {\bf E}}{\partial t}
\cdot\frac{\partial^2 {\bf E}}{\partial t^2}=
-c^2\int_{\Omega\times [0,d]} \frac{\partial {\bf E}}{\partial t}
\cdot {\rm curl}{\rm curl}{\bf E}+ c^2\int_{\Omega\times [0,d]}
\frac{\partial {\bf E}}{\partial t}\cdot\nabla {\rm div}{\bf E}
\end{equation}
where we recalled (\ref{curldel}). Note that ${\rm curl}{\bf E}=0$
on the boundaries $\Omega \times \{ 0 \}$ and $\Omega \times \{ d
\}$ (lower and upper plates). By assuming that ${\bf n}$ is the
outer normal, Green's formulas yield:
$$
\frac{1}{2}\frac{d}{dt}\left[\int_{\Omega\times [0,d]} \left(
\frac{\partial {\bf E}} {\partial t}\right)^{\hspace{-.1cm}2}
~+~c^2 \int_{\Omega\times [0,d]} ({\rm
curl}{\bf E})^2 ~+~c^2\int_{\Omega\times
[0,d]} ({\rm div}{\bf E})^2\right]
$$
$$
~=~c^2\int_{\Omega\times \{ 0\}}   \frac{\partial {\bf E}}
{\partial t} \cdot  {\bf n}~{\rm div}{\bf
E}~+~c^2\int_{\Omega\times \{ d\}} \frac{\partial {\bf E}}
{\partial t} \cdot  {\bf n}~{\rm div}{\bf E}
$$
\begin{equation}\label{wave3}
~+~c^2\int_{\partial\Omega\times [0,d]}   \frac{\partial {\bf E}}
{\partial t} \cdot ({\rm curl}{\bf E}\times {\bf n})~+~
c^2\int_{\partial\Omega\times [0,d]}  \frac{\partial}{\partial
t}\left( {\bf E}\cdot {\bf n}\right) {\rm div}{\bf E}
\end{equation}
All the terms on the right-hand side are zero by virtue of the
homogeneous boundary conditions (note in particular that $\partial
{\bf E}/\partial t=0$ on the lower and upper sides). On the
left-hand side we have the time derivative of a non negative
quantity. Due to the initial conditions, such a quantity is zero
at the beginning. Hence, it will remain zero forever. One easily
finds out that the only compatible solution is ${\bf E}=0$
identically. For this last check, boundary conditions must be
used one more time. Note that the role of the one-dimensional boundaries
$\partial\Omega\times \{ 0\}$ and $\partial\Omega\times \{ d\}$
is negligible (in presence of regular solutions, at least).
\par\smallskip

The above reasoning is quite standard, especially in the framework
of finite element approximations, where the theory is constructed
on a weak formulation having the associated energy similar to the
one considered here.
\par\smallskip

The problem of setting up the correct boundary conditions for the
Maxwell's  first-order system remains open and needs further
discussion. We can guess that ${\bf E}\cdot{\bf n}=0$ and ${\bf
B}\times {\bf n}=0$ are good candidates, because they lead to
reasonable assumptions in the framework of the second-order wave
equation. Nevertheless, we will weaken the second one a little bit.
Concerning with the special case we are discussing, the
electromagnetic energy ``flows'' along characteristic curves contained
in the domain $\Omega\times [0,d]$. Since we do not want {\sl
inflow boundaries} regarding the Poynting field ${\bf P}={\bf E}\times {\bf B}$, 
we can enforce this to be orthogonal to the boundary. We can then replace
(\ref{rote}) by:
\begin{equation}\label{rotem}
{\bf E}\cdot{\bf n}~=~0 ~~~~~~~~~~({\bf E}\times {\bf B})\cdot{\bf n}~=~0
~~~~~{\rm on}~\partial\Omega\times [0,d]~~~~~\forall ~t
\end{equation}
Note that it is not necessary to have ${\bf B}\times {\bf n}=0$ in
order to enforce the milder condition ${\bf P}\cdot{\bf n}=0$. We have plenty of orthogonality
relations: ${\bf E}\perp {\bf n}$,  ${\bf E}\perp {\bf P}$, ${\bf B}\perp {\bf P}$,
 ${\bf P}\perp {\bf n}$, but  ${\bf B}$ remains undertermined.
\par\smallskip

Let us suppose for example that  ${\bf E}$ is forced to be zero on the
lower and upper plates (homogeneous Dirichlet conditions). One
gets, after integrating (\ref{poyn}) in the entire domain:
$$
\frac{1}{2}\frac{d}{d t} \int_{\Omega\times [0,d]}
(\vert {\bf E}\vert^2 +c^2\vert {\bf B}\vert^2)
~=~- c^2 \int_{\Omega\times [0,d]}{\rm div}({\bf E}\times {\bf B})
$$
\begin{equation}\label{poyni}
=- c^2 \int_{\partial\Omega\times [0,d]}({\bf E}\times
{\bf B})\cdot {\bf n}~=~0
\end{equation}
where we used the divergence theorem with ${\bf n}$ directed
outward. The last term is zero because of (\ref{rotem}). According
to (\ref{poyni}), the electromagnetic energy, initially zero, will
stay zero during time evolution. This says that the Maxwell's
system admits unique solution.
\par\smallskip

In alternative to the insulation of the lateral wall, one could
take into account Neumann conditions. Instead of
(\ref{rote}), a viable option is then:
\begin{equation}\label{neum}
\frac{\partial E_x}{\partial {\bf n}}~=~0~~~~~~~~ \frac{\partial
E_y}{\partial {\bf n}}~=~0~~~~~~~~ \frac{\partial E_z}{\partial
{\bf n}}~=~0~~~~~~~~{\rm on}~\partial\Omega\times [0,d]~~~~\forall ~t
\end{equation}
\par\smallskip

By assuming ${\rm div}{\bf E}=0$, we multiply equation (\ref{wave}) by 
$\partial {\bf E}/\partial t$
and integrate, obtaining through the usual passages:
$$
\frac{1}{2}\frac{d}{dt}\left[\int_{\Omega\times [0,d]} \left(
\frac{\partial {\bf E}} {\partial t}\right)^{\hspace{-.1cm}2}
~+~c^2 \int_{\Omega\times [0,d]} (\vert \nabla E_x\vert^2 +
\vert \nabla E_y\vert^2 +\vert \nabla E_z\vert^2)
\right]
$$
\begin{equation}\label{lapl1}
~=~c^2\int_{\partial\Omega\times [0,d]}  \left( 
\frac{\partial E_x}{\partial t}\frac{\partial E_x}{\partial {\bf n}} 
+\frac{\partial E_y}{\partial t}\frac{\partial E_y}{\partial {\bf n}} 
+\frac{\partial E_z}{\partial t}\frac{\partial E_z}{\partial {\bf n}} \right)
~=~0
\end{equation}
where we imposed (\ref{neum}) and homogeneous Dirichlet conditions
on the plates. Again, we deduce an uniqueness result.

\par\medskip

\section{Charging the capacitor}

We apply to a concrete case the situation examined in the previous
section. The two plates, initially short-circuited, are
successively subjected to a difference of potential.
As far as the initial and boundary condition are concerned, we
assume (\ref{initd}) and (\ref{rote}). Moreover, for any $t>0$:
\begin{equation}\label{conv}
{\bf E }~=~(0,0,\alpha (t)) ~~~~~~~{\rm on~ both~plates~}
\end{equation}
where $\alpha$ is a given function with $\alpha (0)=0$ and
$\alpha^\prime (0)=0$. Thus, the charge starts flowing smoothly on
the plates, based on a difference of potential equal to
$V=d\alpha$. For symmetry reasons, we can set the ground at level
$z=d/2$. With respect to the lines of force of the electric field,
for $\alpha$ positive the lower surface is an {\sl inflow
boundary}, while the upper surface is an {\sl outflow boundary}.
The total flux through the boundaries is zero, so that the
integral of ${\rm div}{\bf E}$  on the whole domain is also zero.
We omit to specify the boundary conditions for ${\bf B}$, since
there is no need for them, as it will emerge from the analysis.

%The third component of (\ref{fbm1}) is:
%\begin{equation}\label{fbm13}
%\frac{\partial B_z}{\partial t} ~=~\frac{\partial E_x}{\partial y}
%~-~\frac{\partial E_y}{\partial x}
%\end{equation}
%Thus, $\partial B_z/\partial t=0$ in $S$. Considering that for $t=0$
%one has ${\bf B} =0$, we deduce that $B_z=0$ in $S$, $\forall t$.

\par\smallskip

In the context of smooth functions, the solutions to the Maxwell's
system belong to a subset of those satisfying the wave equation
(\ref{wave}). We can construct an explicit solution by setting
${\bf E}=(0,0,E_z)$, where $E_z$ does not depend on $x$ and $y$.
In this fashion, we have:
\begin{equation}\label{wavez}
\frac{\partial^2 E_z}{\partial t^2}~=~c^2
\frac{\partial^2 E_z}{\partial z^2}~=~c^2 \Delta E_z
\end{equation}
with (see the last relation in (\ref{rotec})):
\begin{equation}\label{condz}
E_z~=~\alpha (t) ~~~{\rm on~ the~plates~}
\end{equation}
\begin{equation}\label{neumz}
 \frac{\partial E_z}{\partial
{\bf n}}~=~0~~~~~~~~{\rm on~the ~lateral~surface}
\end{equation}
\begin{equation}\label{condzi}
E_z~=~0 ~~~~~~~~~ \frac{\partial E_z}{\partial t}~=~0 ~~~~~{\rm at~time~}t=0
\end{equation}
\par\smallskip

The explicit expression of (\ref{wavez}) is not elementary but it
is recoverable through Fourier series expansion. A theory in the
framework of Sobolev spaces can be found for instance in
\cite{evans}, section 7.2, or in \cite{brezis}, p. 345. 
Such a solutions turns out to be smooth in the domain including the 
boundary. Due to the
uniqueness theorem cited in the previous section, there are no
other possible choices for ${\bf E}$. Hence, we also found the solution to the
Maxwell's problem.
\par\smallskip

At this point we notice that the function $E_z$ is not certainly
constant with respect to $z$. In fact, the setting $E_z=\alpha
(t)$, $\forall z \in [0,d]$, is in general incompatible with
equation (\ref{wavez}), because $\partial^2 E_z /\partial t^2
=\alpha^{\prime\prime}(t) \not = 0=\partial^2 E_z /\partial z^2$.
Therefore, the partial derivative $\partial E_z /\partial z$ is
different from zero almost everywhere. In the end, we have:
\begin{equation}\label{nondiv}
{\rm div}{\bf E} ~=~\frac{\partial E_z}{\partial z}   ~\not =~ 0
\end{equation}
\par\smallskip

We just discovered that the solutions to our vector wave equation
cannot be divergenceless. As a consequence, independently on how
we define the magnetic field, there are no chances to solve the
entire set of Maxwell's equations. This is true for any set
$\Omega$, for any $d$ and for any function $\alpha$ (with zero
derivative at the origin); too many degrees of freedom to argue
that this is just incidental.
\par\smallskip

The problem we proposed has very smooth solutions, hence the idea
that things may improve by converting it into variational form is
hopeless. In truth, using a general {\sl test function} in
(\ref{wave2}), a variational formulation is soon obtained, which
can be automatically extended to functions belonging to suitable
Sobolev spaces.
\par\smallskip

By the above procedure, the magnetic field turns out to be totally
unspecified. As already noticed, the electric field remains
parallel to the $z$-axis and does not depend on $x$ and $y$. The
only reasonable choice compatible with such a behavior of the
electric field is ${\bf B}=0$ everywhere at all times. This is
again in contradiction with (\ref{fem1}) since we know that
$\partial{\bf E} /\partial t \not =0$.
\par\smallskip

The case $\alpha^\prime (0)\not =0$ is more difficult to handle,
but this preliminary analysis induces us to guess that conclusions
cannot be too much different. Time periodic conditions applied at
the plates are not compatible with (\ref{initd}). However, even
starting from $\alpha^\prime (0)\not =0$, after a transient, the
function $\alpha$ may assume a given oscillating behavior,
resulting asymptotically in a periodic evolution of the internal
fields (see figure 1).
\par\smallskip

At this point, one may argue that the wave equation approach
requires too strong boundary conditions. Perhaps, by weakening the
insulation condition at the lateral sides one can enlarge the
solution space and obtain situations that are compatible with
${\rm div}{\bf E} =0$. So, we just enforce ${\bf E}\cdot {\bf
n}=0$, forgetting the other constraints and losing the uniqueness
result for the wave equation. Unfortunately, this weaker
hypothesis is again inconsistent with Maxwell's equations. To
prove such a negative claim we follow very classical arguments.
\par\smallskip

We work in the neighborhood of the lower plate $S=\Omega\times \{ 0\}$.
Due to (\ref{conv}), we have:
\begin{equation}\label{derze1}
\frac{\partial E_x}{\partial x} = \frac{\partial E_x}{\partial y} =
\frac{\partial E_y}{\partial x} =\frac{\partial E_y}{\partial y} =
\frac{\partial E_z}{\partial x} =\frac{\partial E_z}{\partial y}=0
~~~~~~~~~{\rm in}~S~~~~ \forall ~t
\end{equation}
\begin{equation}\label{derze}
\frac{\partial^2 E_z}{\partial x^2} =\frac{\partial^2 E_z}{\partial y^2}=0
~~~~~~~~~{\rm in}~S~~~~ \forall ~t
\end{equation}
Thanks to (\ref{derze}), the third component of the wave equation
satisfies (\ref{wavez}) for all $t$, when restricted to $S$.
Moreover, the ${\rm div}{\bf E} =0$ condition implies:
\begin{equation}\label{cv1}
\frac{\partial E_z}{\partial z} ~=~-~\frac{\partial E_x}{\partial x}
~-~\frac{\partial E_y}{\partial y} ~=~0
~~~~~~~~~{\rm in}~S~~~~ \forall ~t
\end{equation}
In conclusion, one obtains:
\begin{equation}\label{cv2}
E_z ~=~\alpha ~~~~~~\frac{\partial E_z}{\partial z}~=~0
~~~~~~\frac{\partial^2 E_z}{\partial z^2}~=~\frac{\alpha^{\prime\prime}}{c^2}
~~~~~~~~~{\rm in}~S~~~~ \forall ~t
\end{equation}
Therefore, near the surface, one gets the Taylor expansion:
\begin{equation}\label{cv3}
E_z ~=~\alpha ~+~\frac{\alpha^{\prime\prime} z^2}{2c^2} ~+~o(z^2)
~~~~~~~~ \forall ~t
\end{equation}
For instance, let us assume that $\alpha$ has a quadratic growth
(though similar considerations will hold for a more general
choice): $\alpha (t)=at^2, ~a>0$. The expression of $E_z$ for $z$
sufficiently small becomes $a(t^2+z^2/c^2)$. We now take $\delta
\in ]0,d]$ and consider the domain $\Omega \times [0,\delta ]$.
The divergence of ${\bf E}$ is zero inside there. The lateral
boundary is insulated, i.e. ${\bf E}\cdot {\bf n}=0$, hence the
incoming flux in $\Omega\times \{0 \}$ must equate the outgoing
flux in $\Omega\times \{\delta \}$. If $\delta$ is suitably small
this is impossible since  ${\bf E}\cdot {\bf n}=at^2$ uniformly in
$\Omega$ for $z=0$, which is less than ${\bf E}\cdot {\bf
n}=a(t^2+\delta^2/c^2)$ for $z=\delta$. Again we arrive at a
contradiction. 
\par\smallskip

It is worthwhile to notice that, in the proof given above,
the zero divergence condition has been recalled both locally
(via (\ref{cv1})) and globally (Gauss's law applied to the
box $\Omega \times [0,\delta ]$).
\par\smallskip

This paradox tells us that there are troubles at the constitutive
level. Relation (\ref{dfem1}) states that condition ${\rm div}{\bf
E}=0$ must be preserved at all times, while we discovered that
this cannot be true. Where is the mistake? The wrong assumption
is in the writing of the Amp\`ere's law in vacuum, where the generic vector
$\partial {\bf E}/\partial t$ is supposed to be the {\sl curl} of
another vector. This is not necessarily true. By dropping this
hypothesis one discovers that, even in absence of currents due to
independent charges, there might be a sort of flow-field with the
property ${\rm div}{\bf E}\not =0$.
\par\smallskip

One may try a correction by rewriting equation (\ref{fem1}) as:
\begin{equation}\label{fem1r}
\frac{\partial {\bf E}}{\partial t}~=~ c^2{\rm curl} {\bf B}~+~(0,0,\alpha^\prime )
\end{equation}
where the added forcing term substitutes the boundary conditions,
that now become of homogeneous type. This
consideration does not help, since the new term is also the {\sl curl}
of a vector (take for instance $(-\frac{1}{2}y\alpha^\prime,
\frac{1}{2}x\alpha^\prime, 0)$).
\par\smallskip

To justify that relation ${\rm div}{\bf E}\not =0$ is physically
admissible one can rely on the finiteness of the speed of light
$c$, which rules the transfer velocity of the information between
the two plates. As one modifies the difference of potential, the
electric field inside the capacitor has to be redistributed
(recall that, in the path we followed, the magnetic field remains equal to zero). 
This happens at speed $c$, in contrast with Coulomb's law (represented
by the Gauss's law) that requires the information to travel
at infinite speed. The only way for a field of the form
$(0,0,E_z)$ to establish a communication between the plates is to
create compression and rarefaction waves by varying its
divergence. Note however that the integral of ${\rm div}{\bf E}$
on the entire domain remains always zero, so that sources and
wells are in perfect equilibrium. These observations, based on
elementary arguments, tell us that a rethinking of the equations
ruling electrodynamics is unavoidable.
\par\smallskip

At this point, some readers may argue that the insulating condition
${\bf E}\cdot {\bf n}=0$ is too ``artificial'' and made on purpose 
to guide the internal field along vertical lines of force.
In order to answer this possible question, first of all, we remark
that, in the framework of 3D geometries, imposing ${\bf E}\cdot {\bf n}=0$
does not necessarily mean that ${\bf E}$ must be vertical (this is
a consequence of the resolution of the wave equation).
Secondly, we propose to remove the boundary conditions in (\ref{rote}) and
replace them by (\ref{neum}). A uniqueness theorem for the wave
equation is still guaranteed (see the end of section 2).
In the new setting, ${\bf E}$ could in principle assume configurations
that are more similar to those of a charged capacitor in a stationary regime,
i.e., with curved lines of force that are more pronounced at the rim
of the plates. By the way, such an improvement is only illusory. Indeed, let us
consider again ${\bf E}=(0,0,E_z)$ (with $E_z$ not depending
on $x$ and $y$) satisfying the one-dimensional equation (\ref{wavez}).
Such a solution also satisfies (\ref{neum}), therefore it is
the unique solution to the vector wave equation (\ref{wave}) equipped
with the new set of boundary constraints. We know that ${\bf E}$ is 
not of Maxwellian type since its divergence is not zero. We also
know that ${\bf B}$ remains equal to zero when time passes.
Thus, starting from ${\bf E}=0$ and $\partial {\bf E}/\partial t=0$
at time $t=0$, there is no development of horizontal component of ${\bf E}$
during the charging process. Such a conclusion seems counterintuitive.
Every physicist would bet that there is a mistake in the reasoning.
On the other hand, this is just a correct mathematical results which is
a consequence of having approached the problem from a nonstandard
perspective. Here the study of a dynamical situation provides
results apparently not in agreement with the stationary case. As claimed in the introduction,
there are many different paths one may follow and they depend
on what information one would like to extrapolate.
Although some partial results could be reasonably in agreement
with the physical phenomenon under study, others may not.
In any case the final answer is often contradictory.
\par\smallskip 

One may finally wonder about the possibility to replace (\ref{neum})
by weaker hypotheses, as done for the case of the insulated wall.
The aim is to allow the creation of a nontrivial field ${\bf B}$
with the consequent generation of curved lines of force for the
electric field. At the same time, the hope is to restore the
zero divergence condition. The risk is to invalid the uniqueness
theorem for Maxwellian solutions. Certainly, one may think that there are infinite
ways the magnetic field can grew up in the capacitor's charging procedure,
as there are infinite ways the electric field may be deformed.
Distinguishing among these solutions may require unusual physical
considerations that are not comprised in the standard theory
of electrodynamics. Nevertheless, such a risk does not occur, since,
even with a total lack of information on the lateral boundary, we
are able to end up with a nonexistence result.
\par\smallskip

The proof is simple.
Let us recall relation  (\ref{cv3}) and for simplicity take $\alpha (t)=at^2$.
Near the surface $S$, the
electric field is of the form: $\Big( E_x, E_y, a(t^2+z^2/c^2)\Big)$, neglecting 
second-order infinitesimals. This is true for any $t$. At level $z=\delta$
(for $\delta >0$ fixed and sufficiently small), by computing the divergence one gets:
\begin{equation}\label{div3}
0~=~{\rm div}{\bf E} ~\approx~\frac{\partial E_x}{\partial x}+\frac{\partial E_y}{\partial y}+
\frac{2a\delta}{c^2}~~~~~~~~ \forall ~t
\end{equation}
where the approximation holds up to first-order infinitesimals. The
crucial step is now to observe that, since the capacitor is initially
uncharged, for $t$ small enough the sum of $\partial E_x/\partial x$ and
$\partial E_y/\partial y$ cannot be equal to the fixed number $-2a\delta /c^2$. Thus, we
arrived at another absurdity.
\par\smallskip

This check shows that there is conflict between boundary constraints,
the wave equation and the divergence-free condition,
pointing out once again the inconsistency of the model.
Note that hypotheses here are really very mild. In particular, it is
sufficient to have the boundary condition ${\bf E}=(0,0,\alpha (t))$
only in a 2D region (of any size) included in $S$. We do not deny that 
there are interesting
solutions of the full set of equations, but they look dotted islands 
in the ocean of electromagnetic phenomena (see the introduction).
\par\smallskip

One can try to ``hide'' the above evidences and argue as follows
(the $n$-steps-instead-of-$n\hspace{-.1cm}+\hspace{-.1cm}1$ \ technique
mentioned in the introduction). Faraday's and Amp\`ere's laws can be advanced in time. 
They have good physical justification, so the solution will naturally find
its path, reproducing within a certain accuracy the phenomenon. Thanks to (\ref{dfem1}),
there is no necessity to check what happens to the divergence
of the electric field. One hopes that nobody will discover that, in some small
remote region of the plates, ${\bf E}$ will grow up in such a way
that $\partial{\bf E}/\partial t$ is not of curl-type, disregarding
Amp\`ere's law. A nonzero divergence starts developing and
at this point the link with Maxwell's equations is definitely lost.
If somebody asks to explain the reason of a non vanishing divergence,
an evasive answer is that such a divergence is negligible in practical
cases. That is why is hard to convince the public of the unreliability
of the Maxwell's model.

\par\medskip

\section{Other paradoxical results}

We propose the following experiment. The difference of potential
between the plates is increased quadratically for a given interval
of time, after which is kept constant. The information propagates from
the boundary to the interior and the third component $E_z$ follows
the wave equation (\ref{wavez}). As we stop the increasing of
potential, the field continues to develop and assumes an
oscillating behavior.
\par\smallskip

The plots of figure 1, obtained from a
simple numerical test, show that after a transient where the
distribution monotonically grows (solid lines), a periodic regime
follows (dashed lines).
%For this test we set: $d=1$, $c=1$, $\alpha (t)=t^2$ for $0\leq t\leq
%\frac{5}{4}$.
We do not specify all the parameters of the test, since the
purpose here is to comment the qualitative behavior. In
particular, the intensity of the field inside the capacitor
assumes values that are greater than the ones attained at the
boundary.
\par\smallskip

By denoting with $A$ the area of $\Omega$, the capacitance
is given by $C=\epsilon_0 A/ d$, while the stored energy is
given by:
\begin{equation}\label{enec}
\frac{1}{2}~CV^2~=~\frac{\epsilon_0}{2}~Ad ~\alpha^2
=~\frac{\epsilon_0}{2}~{\cal V} ~\alpha^2
\end{equation}
where $V=\alpha d$ is the difference of potential and ${\cal V}$
is the measure of the volume of the capacitor. In (\ref{enec}) it
is assumed that, at stationary regime, the electric field inside
the capacitor is uniformly equal to $V/d$. We also know that the
quantity $\frac{1}{2}\epsilon_0 (\vert {\bf E}\vert^2+c^2\vert
{\bf B}\vert^2)$ denotes the energy density of the electromagnetic
field. In the case we are examining, this energy integrated over
the volume of the capacitor is: $\frac{1}{2}\epsilon_0 \int_{\cal
V} E_z^2$. As we notice before, this last quantity can be bigger
than the one predicted by (\ref{enec}).  Of course, perfect
capacitors as the one we are studying here do not exist in reality
and this strangeness is not noticed in practice. Nevertheless,
such a suspicious theoretical result provides us with another
indication that the ruling equation have flaws or, at least, that
the definition of the energy stored by a capacitor is lacunary.
\par\smallskip

\begin{center}
\begin{figure}[h]
\hspace{.8cm}\includegraphics[width=10.cm,height=8.cm]{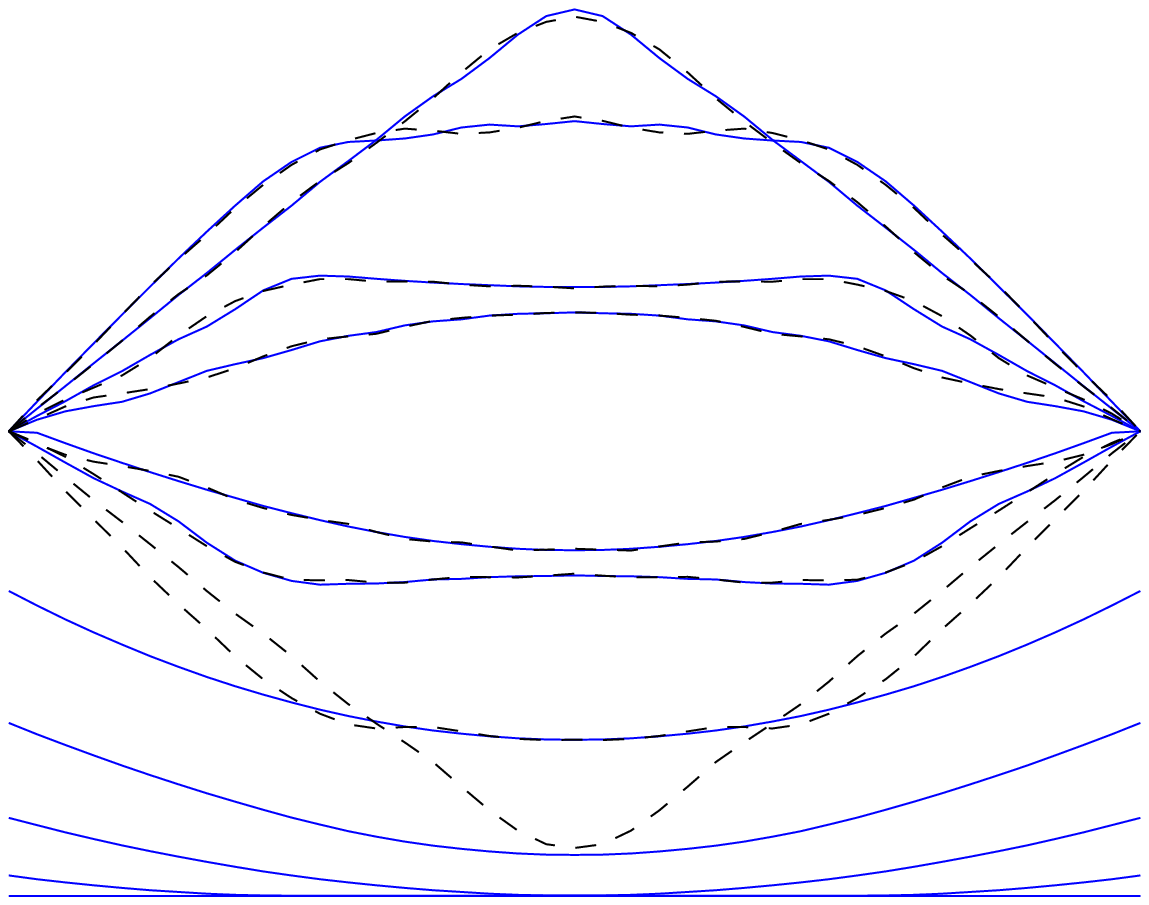}
\vspace{-.1cm}
\begin{caption}{\small Behavior with respect to time of the function
$E_z$. Boundary conditions are increased quadratically and suddenly stopped.
As the wave equation predicts, non dissipating oscillations
are developed.}
\end{caption}
\end{figure}
\end{center}
\vspace{.5cm}

In the {\sl two-capacitors paradox} (see, e.g., \cite{halliday},
vol. 2, p. 684), the charge present at the plates of a capacitor of
capacitance $C$ is redistributed by adding another capacitor in
parallel of capacitance $C$ (the new total capacity is then $2C$).
This implies that the initial voltage $V$ is halved. The initial
stored energy is $\frac{1}{2}CV^2$ while the final one is
$\frac{1}{2}(2C)(V/2)^2=\frac{1}{4}CV^2$. Thus, half of the energy
mysteriously disappears.
\par\smallskip

Explanations of this fact have been proposed copiously. The
principal track for the investigation is the analysis of losses in
the charge transfer procedure. This includes: heating of the
connecting wire, self inductance of the circuit (see
\cite{powell}), electromagnetic emission, etc. What emerges from
this paper is rather the urgent need of restating basic formulas.
The stored energy in (\ref{enec}) is only part of the total energy
that also contains the kinetic of the fields inside the
capacitors. Such a dynamics in inevitably produced by the
redistribution of the charges. By assuming energy conservation,
internal oscillations do not dump and must contribute to the total
energy. A correction of (\ref{enec}) should include this option.
\par\smallskip

A similarity can be made with a perfect elastic ball, that 
initially is at rest at distance $h$ from the ground (pure
gravitational potential energy). The ball successively drops and
bounces back. A barrier is posed at level $h/2$ before the ball
can reach again level $h$. Oscillations develop between the ground
and level $h/2$, where the potential energy is half of the initial
one. However, there is no energy loss, since one has to take into
account the nonzero kinetic energy when the ball hits the upper
obstacle.
\par\smallskip

Formula (\ref{enec}) is only valid in the pure stationary case, so
its application in the dynamical description of the charge
transfer between two capacitors is incorrect. In practical
applications, the oscillations decay in a finite time due to some
internal energy dissipation. This loss of energy must be added to
the one arising from the external circuitry. After an appropriate
interval of time, the total amount
of lost energy is, as correctly predicted, equal to half of the initial 
energy. This result is not as suspicious as before: if from one hand,
we have the right to suppose that the dissipation due to the outer circuit is negligible,
on the other hand we must handle the dissipation of the internal oscillations in some 
appropriate way.
\par\medskip

\section{Comments}

I am sure that following the above discussion many experts will
start providing their explanations. An ideal capacitor does not
exist. Charges are fluctuating on the plates making it impossible
a uniform distribution and resulting in the creation of ``magnetic
currents''. The lateral boundaries of the capacitors cannot be
perfectly insulated. The wave equation theoretically implies that
the divergence of the electric field can be different from zero,
but in a quasi-stationary regime the amount is negligible. For
fast-varying fields the approach to the problem should be
different. At high frequencies, the coupling of electric and
magnetic fields produces electromagnetic emissions. In other
words, nature is very complicated and a simple linear model, such
as the Maxwellian, cannot take into account all the possible
manifestations, unless one accepts to introduce some
approximation.
\par\smallskip

By the way, there are no excuses! What has been studied here is a mathematical 
setting trying to explain a
very simple phenomenon. Moreover, it is not a simple phenomenon
with very rare specific parameters, since the shape of the
capacitor and the applied potential are arbitrary. The results are
wrong because the underlying physics is wrong: too many
constraints compared to the degrees of freedom. The evolution
equations rely on the finiteness of the speed of light, while the
Gauss's law (ruling the stationary cases) finds its roots on the ``action at 
the distance''. Depending on
circumstances, one has to choose what equations are ``more
meaningful''. What is the borderline between the physics of
slow-varying or fast-varying potentials? Nobody can predict it;
because in reality such a difference does not exist. Why do we
think there should be a difference? Because the Maxwell's model is
controversial. It has to be specialized based on the target, and
misses the analysis of the intermediate situations. In fact, very
little is known for instance about the so called {\sl near-field}
of an antenna, where, under suitable resonance conditions,
oscillating fields transform into radiation waves (see my paper
\cite{fun3} to this regard). It is not admissible that the
structure of a model changes depending on the problem to be
solved.
\par\smallskip

I have a solution to propose (see the references to my papers). In
my alternative model equations the divergence of the electric
field can be different from zero even in vacuum (let me skip any
explanation).  What happens inside a capacitor? Everything one may
suspect: magnetic fields are generated, information propagates at
the speed of light, some electromagnetic pressure (included in the
model equations) acts on the plates, and several other known (or
less known) effects. It depends on the assumptions on the device
and the external solicitation. The general solution can be a nightmare. The difficulty of
the math reflects however the complexity of the phenomena observed
in the real world, without any  barrier among the different
regimes. Starting from a unique formulation, negligible quantities
can be simplified later, if the nature of the phenomenon allows
for it. In
the modified model, the wave equation does not hold and it is
substituted by a suitable nonlinear hyperbolic system of equations
where ${\rm div}{\bf E}$ can be actually different from zero. A meaningful
solution (from the mathematical viewpoint) can be finally
recovered. This is the way a correct model should work.
A more precise explanation will be given in a future paper.
\par\smallskip

I still do not know if my proposal is the optimal tool to study
electromagnetism. It is certain that Maxwell's model is
mathematically incorrect and, consequently, requires revision. It
fails on simple questions and not just in the simulation of exotic
problems.

\end{document}